\documentclass[aps]{revtex4}
 \usepackage{amssymb} \usepackage{graphicx}

\begin{document}
 \title{Index of Dirac operators and classification of topological insulators}

\author{\"Umit Ertem}
 \email{umitertemm@gmail.com}
\address{Department of Physics,
Ankara University, Faculty of Sciences, 06100, Tando\u gan-Ankara,
Turkey\\}

\begin{abstract}

Real and complex Clifford bundles and Dirac operators defined on them are considered. By using the index theorems of Dirac operators, table of topological invariants is constructed from the Clifford chessboard. Through the relations between K-theory groups, Grothendieck groups and symmetric spaces, the periodic table of topological insulators and superconductors is obtained. This gives the result that the periodic table of real and complex topological phases is originated from the Clifford chessboard and index theorems.

\end{abstract}

\maketitle

\section{Introduction}

The recent discovery of new topological phases of matter has extended the connections between condensed matter physics and topology \cite{Haldane, Kane Mele, Hasan Kane, Qi Zhang}. The prominent examples of topological phases are topological insulators and they correspond to bulk insulating and edge conducting materials. Topology of the bulk implies the edge conductivity and this differentiates topological insulators from trivial insulators. Besides the classical integer quantum Hall effect, Chern insulators and quantum spin Hall effect are also realized experimentally as examples of topological insulators \cite{Bernevig Hughes Zhang, Konig et al}. Moreover, topological superconductors can also be described as bulk insulating and edge conducting materials that exemplifies the topological phases of matter \cite{Qi Zhang}. This bulk/edge correspondence in topological materials is a manifestation of the holography principle in condensed matter physics. The existence of topological phases depends on the symmetries of the Hamiltonian and the time-reversal (T) and charge conjugation (C) symmetries play prominent roles in the classification of topological materials. Topological insulators and superconductors can be characterized by the relevant topological invariants defined from the eigenstates of the corresponding Hamiltonians. There are two types of topological materials depending on the group that the topological invariants take values which are $\mathbb{Z}$-insulators (Chern insulators) and $\mathbb{Z}_2$-insulators. $\mathbb{Z}$-invariants are generally characterized by Chern numbers or winding numbers defined from the Berry curvature of the system and $\mathbb{Z}_2$-invariants correspond to Kane-Mele invariants or Chern-Simons invariants \cite{Kane Mele2, Wang Qi Zhang}.

Classification of topological phases in different dimensions and symmetry classes has been obtained by using Clifford algebras and K-theory \cite{Kitaev, Ryu Schnyder Furusaki Ludwig, Stone Chiu Roy, Abramovici Kalugin, Freed Moore, Budich Trauzettel, Kaufmann et al}. Because of the Bott periodicity of K-groups, the topological phases can be described by a finite periodic table. The rows of the table correspond to the ten different symmetry classes of Hamiltonians which are classified in \cite{Altland Zirnbauer} and the columns describe different dimensions. The periodic table of topological insulators and superconductors shows that there are five different topological symmetry classes in each dimension and three of them are $\mathbb{Z}$-insulators and two of them are $\mathbb{Z}_2$-insulators. Since the symmetry classes of Hamiltonians can be related to Cartan symmetric spaces, the periodic table can be divided into two parts that characterize the complex and real classes separately. Besides Clifford algebras and K-theory, the periodic table can also be obtained by analyzing the stability of gapless boundary states against perturbations or by the dimensional reduction method of massive Dirac Hamiltonians \cite{Ryu Schnyder Furusaki Ludwig}. Although the relations between Clifford algebras, K-groups and topological phases are known, the origin of the periodic table in mathematical terms is not discussed extensively in the literature.

In this paper, we show that the periodic table of topological insulators and superconductors can be obtained directly from the Clifford chessboard of Clifford algebras and the index theorems of Dirac operators. We start by turning the Clifford chessboard to the table of Clifford bundles and defining relevant Dirac operators on these bundles. Index theorems for Dirac operators relates the analytic index to the topological index \cite{Atiyah Singer, Atiyah Singer2, Atiyah} and we find the table of indices of Dirac operators in that way. From the relations between index theorems, K-theory groups, Grothendieck groups of Clifford algebra representations and symmetric spaces, we obtain the periodic table of topological phases for all dimensions and symmetry classes. These steps are applied both to real and complex classes separately. Instead of starting with the symmetry classes of free-fermion Hamiltonians and obtaining the topological classes by using Clifford algebras and K-theory which is done in the literature \cite{Kitaev, Stone Chiu Roy, Abramovici Kalugin, Morimoto Furusaki}, we start from the Clifford chessboard and find the symmetry classes of Hamiltonians and topological phases by using the index theorems of Dirac operators. This approach gives the different topological classes in all dimensions and relates the topological invariants in these classes with the indices of Dirac operators.

The paper is organized as follows. In Section 2, the mathematical concepts discussed in the subsequent sections are connected to the properties of the physical systems that describe topological materials. Section 3 includes the periodicity relations of Clifford algebras and the construction of the Clifford chessboard. In Section 4, we obtain the periodic table of real topological phases by starting with the table of Clifford bundles and by using the index theorems of Dirac operators and relations with K-groups, Grothendieck groups and symmetric spaces. Section 5 deals with the problem of obtaining the periodic table of complex topological phases with the methods discussed in Section 4. Section 6 concludes the paper.

\section{Dirac Hamiltonians of Topological Materials}

In this section, the mathematical concepts used in the subsequent sections to obtain the classification table of topological insulators are connected to the properties of real physical systems via their Hamiltonians and eigenstates. The free-fermion Hamiltonians are classified in ten different symmetry classes depending on the existence or non-existence of anti-unitary symmetries such as time-reversal and charge-conjugation \cite{Altland Zirnbauer}. These so-called Cartan classes represent the defining Hamiltonians of free-fermion systems. The low energy effective Hamiltonians of topological insulators and superconductors in all dimensions and symmetry classes are characterized by Dirac-like Hamiltonians. They can be written generally in terms of the momentum variables $\bf{k}$ in the following form \cite{Fruchart Carpentier,Kaufmann et al}
\begin{equation}
H(\bf{k})=\bf{d}(\bf{k}).\bf{\sigma}
\end{equation}
where $\bf{d}(\bf{k})$ denotes the functions written in terms of $\bf{k}$ and $\bf{\sigma}$ are Clifford algebra generators in relevant dimensions. For example, the two dimensional Haldane model of graphene \cite{Haldane}, which is a Chern insulator characterized by a $\mathbb{Z}$ topological invariant, has the following Bloch Hamiltonian
\begin{eqnarray}
H({\bf{k}})=\sum_{i=1}^3\left\{2t_2\cos(\phi)\cos({\bf{k}}.{\bf{b}}_i)\sigma_0+t_1\left[\cos({\bf{k}}.{\bf{a_i}})\sigma_1+\sin({\bf{k}}.{\bf{a}}_i)\sigma_2\right]+\left[\frac{M}{3}-2t_2\sin(\phi)\sin({\bf{k}}.{\bf{b}}_i)\right]\sigma_3\right\}
\end{eqnarray}
where $t_1$ and $t_2$ are nearest neighbour and next nearest neighbour hopping parameters, $\phi$ is the Haldane phase and ${\bf{a}}_i$ and ${\bf{b}}_i$ are nearest neighbour and next nearest neighbour displacement vectors. $M$ is the on-site energy and $\sigma_i$ are Pauli matrices that satisfy the Clifford algebra relations. At the low energy limit around the ${\bf{K}}$ point, this Hamiltonian reduces to the following continuum limit
\begin{eqnarray}
H({\bf{K}})=-3t_2\cos(\phi)\sigma_0+\frac{3}{2}t_1\left(\kappa_2\sigma_1-\kappa_1\sigma_2\right)+\left(M-3\sqrt{3}t_2\sin(\phi)\right)\sigma_3
\end{eqnarray}
where $\kappa_i$ are the momentum space Pauli matrices. This Hamiltonian is in the form of a Dirac Hamiltonian which corresponds to a Dirac operator written in terms of the Clifford algebra basis $\sigma_i$. Similarly, the two-dimensional time-reversal invariant $\mathbb{Z}_2$ insulator which is characterized by the Kane-Mele model \cite{Kane Mele} has the following form of Bloch Hamiltonian
\begin{eqnarray}
H({\bf{k}})=\sum_{i=1}^5d_i({\bf{k}})\Gamma_i+\sum_{i<j=1}^5d_{ij}({\bf{k}})\Gamma_{ij}
\end{eqnarray}
where $\Gamma_i$ and $\Gamma_{ij}=\frac{1}{2i}[\Gamma_i,\Gamma_j]$ are Clifford algebra generators which are defined in terms of sublattice Pauli matrices $\sigma_i$ and spin Pauli matrices $s_i$ as follows
\begin{eqnarray}
\Gamma_{1,2,3,4,5}=\left(\sigma_1\otimes s_0, \sigma_3\otimes s_0, \sigma_2\otimes s_1, \sigma_2\otimes s_2, \sigma_2\otimes s_3\right).
\end{eqnarray}
The functions $d_i({\bf{k}})$ and $d_{ij}({\bf{k}})$ are given by
\begin{eqnarray}
d_1({\bf{k}})&=&t\left[1+2\cos\left(\frac{k_x}{2}\right)\cos\left(\frac{\sqrt{3}k_y}{2}\right)\right]\nonumber\\
d_2({\bf{k}})&=&M\nonumber\\
d_3({\bf{k}})&=&\lambda_R\left[1-\cos\left(\frac{k_x}{2}\right)\cos\left(\frac{\sqrt{3}k_y}{2}\right)\right]\nonumber\\
d_4({\bf{k}})&=&-\sqrt{3}\lambda_R\sin\left(\frac{k_x}{2}\right)\sin\left(\frac{\sqrt{3}k_y}{2}\right)\\
d_{12}({\bf{k}})&=&-2t\cos\left(\frac{k_x}{2}\right)\sin\left(\frac{\sqrt{3}k_y}{2}\right)\nonumber\\
d_{15}({\bf{k}})&=&\lambda_{SO}\left[2\sin(k_x)-4\sin\left(\frac{k_x}{2}\right)\cos\left(\frac{\sqrt{3}k_y}{2}\right)\right]\nonumber\\
d_{23}({\bf{k}})&=&-\lambda_R\cos\left(\frac{k_x}{2}\right)\sin\left(\frac{\sqrt{3}k_y}{2}\right)\nonumber\\
d_{24}({\bf{k}})&=&\sqrt{3}\lambda_R\sin\left(\frac{k_x}{2}\right)\cos\left(\frac{\sqrt{3}k_y}{2}\right)\nonumber
\end{eqnarray}
where $t$ is the nearest neighbour hopping parameter and $\lambda_{SO}$ and $\lambda_R$ are spin-orbit and Rashba coupling parameters, respectively. One can easily see that the continuum limit of Hamiltonian (4) corresponds to a sum of two Haldane model Dirac Hamiltonians with different spins. So, we again have a corresponding Dirac operator written in terms of the Clifford algebra basis given in (5). It is also known that the three-dimensional models that describe topological materials again have effective Dirac Hamiltonians written in terms of relevant Clifford algebra basis. Indeed, effective Dirac Hamiltonians are characteristic properties of physical systems that correspond to topological insulators and superconductors in all dimensions \cite{Shen Shan Lu}. Although the real physical systems arise in two and three dimensions, the mathematical structures describing Dirac Hamiltonians and topological materials can be generalized to all higher dimensional systems. Depending on the symmetry classes of physical models determined by the existence or non-existence of anti-unitary symmetries such as time-reversal, charge conjugation and chiral symmetries, the Clifford algebra generators appearing in the effective Dirac Hamiltonians can be positive or negative generators of the algebra. So, the symmetry properties of the Hamiltonians determine the properties of the Clifford algebra generators. On the other hand, the eight-fold periodicity of real Clifford algebras and the two-fold periodicity of complex Clifford algebras which are given in (16) simplify the dimensional pattern of Dirac Hamiltonians for higher dimensional topological materials. This means that in real class topological materials, the symmetry properties of Dirac Hamiltonians for different dimensions and symmetry classes repeat itself after eight dimensions. Similarly, for complex class topological materials, the properties of Dirac Hamiltonians repeat itself after two dimensions. Namely, the properties of all higher dimensional models of topological materials can be deduced from the lower dimensional systems.

In condensed matter systems, the electronic states are described by Bloch wave functions. So, the eigenvalue equation for the above Hamiltonians can be written as
\begin{equation}
H({\bf{k}})u_n({\bf{k}})=E_n({\bf{k}})u_n({\bf{k}})
\end{equation}
where $E_n(\bf{k})$ are $n$ eigenvalues corresponding to $n$ eigenstates $u_n(\bf{k})$. Indeed, $u_n(\bf{k})$ correspond to the sections of a rank $n$ Hilbert bundle $E\rightarrow B$ which is called Bloch bundle over the Brillouin zone $B$ of the system. Since the Hamiltonians are constructed from the Clifford algebra generators, $u_n(\bf{k})$ are elements of the representation space of Clifford algebras. Thus, the bundle $E$ corresponding to the bundle of eigenstates of the Hamiltonian is in fact a Dirac bundle which is a bundle corresponding to a left module of the Clifford algebra as defined in Section 4.A and the Dirac Hamiltonians of topological materials correspond to the Dirac operators on this bundle. Moreover, the anti-unitary symmetries that define the symmetry classes of Dirac Hamiltonians play the role of $Cl_k$-actions that are defined in Section 4.A. So, the Dirac bundles and Dirac operators defined in sections 4.A and 4.B correspond to these physical structures defined for topological materials in real classes and similarly the Dirac operators defined in section 5 correspond to complex class topological materials.

Index structure of Dirac-like Hamiltonians that appear in topological materials is compatible with the index theorems of Dirac operators. For some special cases in low dimensions, it is shown that indices of Dirac-like Hamiltonians correspond to topological invariants that characterize non-trivial topological classes of topological insulators and superconductors \cite{Fukui Shiozaki Fujiwara Fujimoto}. The analysis in section 4.C gives a general relation between the indices of Dirac operators corresponding to Dirac Hamiltonians of topological materials and the topological invariants of topological insulators and superconductors in all dimensions and symmetry classes. $\mathbb{Z}$-class topological invariants which are Chern and winding numbers and $\mathbb{Z}_2$-class topological invariants that are Kane-Mele and Chern-Simons invariants naturally arise from the indices of Dirac operators defined in (22) and (29). So, the analysis gives correct periodic table of topological materials and shows that the topological invariants characterizing the topological classes come from the indices of the representing Dirac operators in all cases. In addition, it also shows that the classification and periodic table of topological materials are originated from the Clifford algebras that are used in the construction of Dirac Hamiltonians of these materials.

Adding energy levels which do not cross the Fermi energy to the system defined in (1) does not give a non-trivial topological structure. Namely, adding topologically trivial bands to the Bloch bundle does not change the topological class of the system. So, for the Bloch bundle $E$, we have an equivalence of bundles
\begin{equation}
E\oplus F^n\cong E\oplus G^m
\end{equation}
where $F^n$ and $G^m$ are $n$-dimensional and $m$-dimensional trivial bundles, respectively. This is the definition of the stable equivalence of bundles and stably equivalent bundles are classified by K-theory. So, $KO$-groups and $K$-groups appearing in the analysis of sections 4.D and 5 are related to the Bloch bundles defined on the Brillouin zones of topological materials. They show the stable equivalence classes of these Bloch bundles in different dimensions and symmetry classes. However, there are two different cases for the Brillouin zones of topological materials; periodic lattices and continuum models. The Brillouin zones of periodic lattices are described by $d$-dimensional torus $T^d$ and for the continuum models they correspond to $d$-dimensional spheres $S^d$ \cite{Budich Trauzettel}. Continuum models are effective low energy large distance theories and as is stated above that the Dirac-like Hamiltonians are low energy effective Hamiltonians of topological materials. For large $\bf{k}$, the energy bands generally have a trivial topological structure and by taking a one-point compactification at $\bf{k}\rightarrow\infty$, the Brillouin zone of the model has the topology of $S^d$. Since the analysis of K-theory groups in the paper is based on the bundles over $S^d$, they only correspond to the continuum models of topological materials. Classification scheme for periodic lattices is more complicated and may give a different periodic table from the one in the paper since the calculation of K-theory groups of $T^d$ can give different results than for $S^d$. Indeed, the K-groups of $T^d$ can be written as direct sums of K-groups of $S^d$ as is given in \cite{Kitaev} and they characterize the so-called weak topological insulators. On the other hand, the continuum models and K-groups of $S^d$ determine the topological classes of strong topological insulators.

\section{Clifford algebras and periodicity relations}

On a vector space $V$ with a quadratic form $q$, the Clifford algebra $Cl(V,q)$ is generated by $V$ with the multiplication rule
\begin{equation}
x.y+y.x=-2q(x,y)
\end{equation}
for $x,y\in V$. If we take the vector space $V=\mathbb{R}^{p+q}$ with the quadratic form $Q(x)=q(x,x)$ which is given by
\begin{equation}
Q(x)=x_1^2+...+x_p^2-x_{p+1}^2-...-x_{p+q}^2
\end{equation}
then we denote the Clifford algebra on $\mathbb{R}^{p+q}$ as $Cl_{p,q}\equiv Cl(\mathbb{R}^{p+q},Q)$ and it is generated by $p$ negative and $q$ positive generators. In particular, if we consider any $Q$-orthonormal bases $\{e_1,...e_{p+q}\}$ of $\mathbb{R}^{p+q}$, then $Cl_{p,q}$ is generated by $e_1,..., e_{p+q}$ that satisfy the following relation
\begin{eqnarray}
e_i.e_j+e_j.e_i&=&\left\{
                                                                               \begin{array}{ll}
                                                                                 -2\delta_{ij}, & \hbox{ for $i\leq p$} \\
                                                                                 +2\delta_{ij}, & \hbox{ for $i>p$}
                                                                               \end{array}
                                                                             \right.
\end{eqnarray}
The dimension of the Clifford algebra $Cl_{p,q}$ for $p+q=n$ is equal to $2^n$. As special cases, we denote
\begin{eqnarray}
Cl_n\equiv Cl_{n,0}\nonumber\\
Cl^*_n\equiv Cl_{0,n}
\end{eqnarray}
for the Clifford algebras with only negative and only positive generators, respectively. An automorphism $\eta:Cl_{p,q}\rightarrow Cl_{p,q}$ can be defined on $Cl_{p,q}$ that gives the $\mathbb{Z}_2$-grading of the Clifford algebra; $Cl_{p,q}=Cl^0_{p,q}\oplus Cl^1_{p,q}$. This corresponds to the superalgebra structure of the Clifford algebras and the even part $Cl^0_{p,q}$ constitute a subalgebra structure. There is an algebra isomorphism between Clifford algebras and their even subalgebras which is written as
\begin{eqnarray}
Cl_{p,q}\cong Cl^0_{p+1,q}\nonumber\\
Cl_n\cong Cl^0_{n+1}
\end{eqnarray}
Other isomorphisms for Clifford algebras are \cite{Lawson Michelsohn}
\begin{eqnarray}
Cl_{p,q}\cong Cl_{p-4,q+4}\nonumber\\
Cl_{p,q+1}\cong Cl_{q,p+1}
\end{eqnarray}
Moreover, some tensor product isomorphisms of Clifford algebras can also be stated as follows \cite{Lawson Michelsohn}
\begin{eqnarray}
Cl_{p,q+2}\cong Cl_{q,p}\otimes Cl_{0,2}\nonumber\\
Cl_{p+2,q}\cong Cl_{q,p}\otimes Cl_{2,0}\\
Cl_{p+1,q+1}\cong Cl_{p,q}\otimes Cl_{1,1}\nonumber
\end{eqnarray}

On the other hand, if we choose $V=\mathbb{C}^{p+q}$, then we can define the complex Clifford algebras as the complexification of real Clifford algebras with complex quadratic form; $\mathbb{C}l_{p,q}\equiv Cl_{p,q}\otimes_{\mathbb{R}}\mathbb{C}\cong Cl(\mathbb{C}^{p+q},Q\otimes\mathbb{C})$. Since there is no distinction between positive and negative generators in the complex case, we can simply denote the complex Clifford algebras as $\mathbb{C}l_n$. The structure of real and complex Clifford algebras can be obtained from the following periodicity relations \cite{Lawson Michelsohn, Benn Tucker}
\begin{eqnarray}
Cl_{p+8,q}&\cong& Cl_{p,q}\otimes Cl_{8,0}\nonumber\\
Cl_{p,q+8}&\cong& Cl_{p,q}\otimes Cl_{0,8}\\
\mathbb{C}l_{n+2}&\cong& \mathbb{C}l_n\otimes_{\mathbb{C}}\mathbb{C}l_2\nonumber
\end{eqnarray}
As can be seen from the above isomorphisms, while the real Clifford algebras have eight-fold periodicity, the complex Clifford algebras have two-fold periodicity. Indeed, Clifford algebras correspond to simple or semi-simple matrix algebras constructed from the division algebras $\mathbb{R}$, $\mathbb{C}$ and $\mathbb{H}$. From a simple observation of the Clifford algebra relation (11) satisfied by the generators of the Clifford algebra, the lower dimensional Clifford algebras can be obtained in terms of division algebras as follows
\begin{eqnarray}
Cl_{1,0}&\cong& \mathbb{C}\nonumber\\
Cl_{0,1}&\cong& \mathbb{R}\oplus\mathbb{R}\nonumber\\
Cl_{1,1}&\cong& \mathbb{R}(2)\nonumber\\
Cl_{2,0}&\cong& \mathbb{H}\nonumber\\
Cl_{0,2}&\cong& \mathbb{R}(2)\nonumber
\end{eqnarray}
where $\mathbb{R}(2)$ denotes the 2$\times$2 matrix algebra with real elements. These basic Clifford algebras and the periodicity relations in (14), (15) and (16) determine the corresponding division algebras of higher dimensional Clifford algebras as in the following table \cite{Benn Tucker}

\quad\\
{\centering{
\begin{tabular}{c c}

% after \\: \hline or \cline{col1-col2}
\hline \hline
$q-p(\text{mod }8)$ & $Cl_{p,q}$ \\ \hline
$0 , 2$ & $\mathbb{R}(2^{(p+q)/2})$ \\
$3 , 7$ & $\mathbb{C}(2^{(p+q-1)/2})$ \\
$4 , 6$ & $\mathbb{H}(2^{(p+q-2)/2})$ \\
$1$ & $\mathbb{R}(2^{(p+q-1)/2})\oplus\mathbb{R}(2^{(p+q-1)/2})$ \\
$5$ & $\mathbb{H}(2^{(p+q-3)/2})\oplus\mathbb{H}(2^{(p+q-3)/2})$ \\ \hline \hline

\end{tabular}
\quad\\
\quad\\
\quad\\

\centering
\begin{tabular}{c c}

% after \\: \hline or \cline{col1-col2}
\hline \hline
$n (\text{mod }2)$ & $\mathbb{C}l_{n}$ \\ \hline
$\text{0}$ & $\mathbb{C}(2^{n/2})$ \\
$\text{1}$ & $\mathbb{C}(2^{(n-1)/2})\oplus\mathbb{C}(2^{(n-1)/2})$ \\ \hline \hline

\end{tabular}
\quad\\
\quad\\
\quad\\}}
where $\mathbb{R}(2^n)$, $\mathbb{C}(2^n)$ and $\mathbb{H}(2^n)$ denotes the $2^n\times 2^n$ matrix algebras that take values in $\mathbb{R}$, $\mathbb{C}$ and $\mathbb{H}$, respectively. Especially, we can write the Clifford algebras $Cl^*_n$ and $\mathbb{C}l_n$ as follows

\quad\\
{\centering{
\begin{tabular}{c c c c c c c c c}

% after \\: \hline or \cline{col1-col2} ...
\hline \hline
$n$ & $0$ & $1$ & $2$ & $3$ & $4$ & $5$ & $6$ & $7$ \\ \hline
$Cl^*_n$ & \,\,\,\,$\mathbb{R}$\,\,\,\, & \,\,\,\,$\mathbb{R}\oplus\mathbb{R}$\,\,\,\, & \,\,\,\,$\mathbb{R}(2)$\,\,\,\, & \,\,\,\,$\mathbb{C}(2)$\,\,\,\, & \,\,\,\,$\mathbb{H}(2)$\,\,\,\, & \,\,\,\,$\mathbb{H}(2)\oplus\mathbb{H}(2)$\,\,\,\, & \,\,\,\,$\mathbb{H}(4)$\,\,\,\, & \,\,\,\,$\mathbb{C}(8)$\,\,\,\, \\
$\mathbb{C}l_n$ & \,\,\,\,$\mathbb{C}$\,\,\,\, & \,\,\,\,$\mathbb{C}\oplus\mathbb{C}$\,\,\,\, & \,\,\,\,$\mathbb{C}(2)$\,\,\,\, & \,\,\,\,$\mathbb{C}(2)\oplus\mathbb{C}(2)$\,\,\,\, & \,\,\,\,$\mathbb{C}(4)$\,\,\,\, & \,\,\,\,$\mathbb{C}(4)\oplus\mathbb{C}(4)$\,\,\,\, & \,\,\,\,$\mathbb{C}(8)$\,\,\,\, & \,\,\,\,$\mathbb{C}(8)\oplus\mathbb{C}(8)$\,\,\,\, \\ \hline \hline

\end{tabular}}
\quad\\
\quad\\
\quad\\}

From the considerations given above, one can construct a table of Clifford algebras which is called the Clifford chessboard in the following way

\quad\\
{\centering{
\begin{tabular}{c c c c c c c c c}

% after \\: \hline or \cline{col1-col2} ...
\hline \hline
$Cl_{n,s}$ & $n=0$ & $1$ & $2$ & $3$ & $4$ & $5$ & $6$ & $7$ \\ \hline
$s=0$ & $\mathbb{R}$ & $\mathbb{C}$ & $\mathbb{H}$ & $\mathbb{H}\oplus\mathbb{H}$ & $\mathbb{H}(2)$ & $\mathbb{C}(4)$ & $\mathbb{R}(8)$ & $\mathbb{R}(8)\oplus\mathbb{R}(8)$ \\
$1$ & $\mathbb{R}\oplus\mathbb{R}$ & $\mathbb{R}(2)$ & $\mathbb{C}(2)$ & $\mathbb{H}(2)$ & $\mathbb{H}(2)\oplus\mathbb{H}(2)$ & $\mathbb{H}(4)$ & $\mathbb{C}(8)$ & $\mathbb{R}(16)$ \\
$2$ & $\mathbb{R}(2)$ & $\mathbb{R}(2)\oplus\mathbb{R}(2)$ & $\mathbb{R}(4)$ & $\mathbb{C}(4)$ & $\mathbb{H}(4)$ & $\mathbb{H}(4)\oplus\mathbb{H}(4)$ & $\mathbb{H}(8)$ & $\mathbb{C}(16)$ \\
$3$ & $\mathbb{C}(2)$ & $\mathbb{R}(4)$ & $\mathbb{R}(4)\oplus\mathbb{R}(4)$ & $\mathbb{R}(8)$ & $\mathbb{C}(8)$ & $\mathbb{H}(8)$ & $\mathbb{H}(8)\oplus\mathbb{H}(8)$ & $\mathbb{H}(16)$ \\
$4$ & $\mathbb{H}(2)$ & $\mathbb{C}(4)$ & $\mathbb{R}(8)$ & $\mathbb{R}(8)\oplus\mathbb{R}(8)$ & $\mathbb{R}(16)$ & $\mathbb{C}(16)$ & $\mathbb{H}(16)$ & $\mathbb{H}(16)\oplus\mathbb{H}(16)$ \\
$5$ & $\mathbb{H}(2)\oplus\mathbb{H}(2)$ & $\mathbb{H}(4)$ & $\mathbb{C}(8)$ & $\mathbb{R}(16)$ & $\mathbb{R}(16)\oplus\mathbb{R}(16)$ & $\mathbb{R}(32)$ & $\mathbb{C}(32)$ & $\mathbb{H}(32)$ \\
$6$ & $\mathbb{H}(4)$ & $\mathbb{H}(4)\oplus\mathbb{H}(4)$ & $\mathbb{H}(8)$ & $\mathbb{C}(16)$ & $\mathbb{R}(32)$ & $\mathbb{R}(32)\oplus\mathbb{R}(32)$ & $\mathbb{R}(64)$ & $\mathbb{C}(64)$ \\
$7$ & $\mathbb{C}(8)$ & $\mathbb{H}(8)$ & $\mathbb{H}(8)\oplus\mathbb{H}(8)$ & $\mathbb{H}(16)$ & $\mathbb{C}(32)$ & $\mathbb{R}(64)$ & $\mathbb{R}(64)\oplus\mathbb{R}(64)$ & $\mathbb{R}(128)$ \\ \hline \hline

\end{tabular}}
\quad\\
\quad\\
\quad\\}

\section{Real Clifford bundles and real classes of periodic table}

In this section, we consider vector bundles whose fibres correspond to the real Clifford algebras. By considering Dirac operators and index theorems, we will obtain the periodic table of real classes in the classification of topological insulators and superconductors in several steps.

\subsection{$Cl_k$-bundles and table of $Cl^*_{s-n}$}

On a manifold $M$, the Clifford bundle $Cl(E)$ is defined as the bundle of Clifford algebras over $M$. If the fibers correspond to the real Clifford algebras, then we call it the real Clifford bundle. The algebra structure of the space of sections of $Cl(E)$ arises from the fibrewise multiplication in $Cl(E)$. Moreover, a Dirac bundle over $M$ is a bundle $S$ of left modules of $Cl(M)$ with a Riemannian metric and a connection on it where $Cl(M)$ is the Clifford algebra defined on $M$. A special case of a Dirac bundle is the spinor bundle over $M$ whose fibres correspond to the spinor spaces as left modules of $Cl(M)$.

The Clifford algebra $Cl_k$ denotes the real Clifford algebra with
$k$ negative generators as defined in Section 3. A $Cl_k$-Dirac
bundle over $M$ is a real Dirac bundle $S$ over $M$ with a right
action $Cl_k\hookrightarrow\text{Aut}(S)$ which is parallel and
commutes with multiplication by the elements of $Cl(M)$. So, a $Cl_k$-Dirac bundle is a Dirac bundle with an extra multiplication by $Cl_k$. Because of the $\mathbb{Z}_2$-grading property of $Cl_k$, the $Cl_k$-Dirac bundle also carry a $\mathbb{Z}_2$-grading $S=S^0\oplus S^1$ with a $\mathbb{Z}_2$-grading for the $Cl_k$-action. Similarly, one can define $Cl^*_k$ -Dirac bundles from the definition (4) whose fibres correspond to the left modules of Clifford algebras with $k$ positive generators.

Let us consider $Cl^*_k$-Dirac bundles. Since the real Clifford algebras have eight-fold periodicity as stated in (16), it is enough to consider $Cl^*_{k(\text{mod }8)}$. We can construct a square table of $8\times 8$ entries whose elements correspond to $Cl^*_{s-n (\text{mod }8)}$ -Dirac bundles where $s,n=0,1,...,7$

\quad\\
{\centering{
\begin{tabular}{c c c c c c c c c}

% after \\: \hline or \cline{col1-col2} ...
\hline \hline
$Cl^*_{s-n (\text{mod }8)}$ & $n=0$ & $1$ & $2$ & $3$ & $4$ & $5$ & $6$ & $7$ \\ \hline
$s=0$ & $Cl^*_0$ & $Cl^*_7$ & $Cl^*_6$ & $Cl^*_5$ & $Cl^*_4$ & $Cl^*_3$ & $Cl^*_2$ & $Cl^*_1$ \\
$1$ & $Cl^*_1$ & $Cl^*_0$ & $Cl^*_7$ & $Cl^*_6$ & $Cl^*_5$ & $Cl^*_4$ & $Cl^*_3$ & $Cl^*_2$ \\
$2$ & $Cl^*_2$ & $Cl^*_1$ & $Cl^*_0$ & $Cl^*_7$ & $Cl^*_6$ & $Cl^*_5$ & $Cl^*_4$ & $Cl^*_3$ \\
$3$ & $Cl^*_3$ & $Cl^*_2$ & $Cl^*_1$ & $Cl^*_0$ & $Cl^*_7$ & $Cl^*_6$ & $Cl^*_5$ & $Cl^*_4$ \\
$4$ & $Cl^*_4$ & $Cl^*_3$ & $Cl^*_2$ & $Cl^*_1$ & $Cl^*_0$ & $Cl^*_7$ & $Cl^*_6$ & $Cl^*_5$ \\
$5$ & $Cl^*_5$ & $Cl^*_4$ & $Cl^*_3$ & $Cl^*_2$ & $Cl^*_1$ & $Cl^*_0$ & $Cl^*_7$ & $Cl^*_6$ \\
$6$ & $Cl^*_6$ & $Cl^*_5$ & $Cl^*_4$ & $Cl^*_3$ & $Cl^*_2$ & $Cl^*_1$ & $Cl^*_0$ & $Cl^*_7$ \\
$7$ & $Cl^*_7$ & $Cl^*_6$ & $Cl^*_5$ & $Cl^*_4$ & $Cl^*_3$ & $Cl^*_2$ & $Cl^*_1$ & $Cl^*_0$ \\ \hline \hline

\end{tabular}}
\quad\\
\quad\\
\quad\\}

The table can be extended to bigger values of $s$ and $n$, however, it repeats the pattern because of the eight-fold periodicity. Moreover, it is equivalent to the table of Clifford chessboard because of the following isomorphism
\[
Cl_{n,s}\cong Cl^*_{s-n(\text{mod }8)}.
\]
So, we obtain the table of $Cl^*_{s-n (\text{mod }8)}$ -Dirac bundles from the Clifford chessboard of Clifford algebras.

\subsection{$Cl_k$-Dirac operators}

On a Dirac bundle $S$ on $M$ with sections $\Gamma(S)$, a first-order differential operator $D:\Gamma(S)\longrightarrow\Gamma(S)$ called Dirac operator can be defined. For the frame basis $\{X_a\}$ and co-frame basis $\{e^a\}$, the Dirac operator can be written in terms of the connection $\nabla$ and local coordinates as
\begin{equation}
D=e^a.\nabla_{X_a}
\end{equation}
where $.$ denotes the Clifford product. On a Riemannian manifold $M$, the Dirac operator of any Dirac bundle is self-adjoint. If the Dirac bundle is $\mathbb{Z}_2$-graded $S=S^0\oplus S^1$, then the Dirac operator can be written in the form
\begin{equation}
D=\left(\begin{array}{cc}
0 & D^1 \\
D^0 & 0 \\
\end{array}\right)
\end{equation}
where $D^0:\Gamma(S^0)\longrightarrow\Gamma(S^1)$ and $D^1:\Gamma(S^1)\longrightarrow\Gamma(S^0)$. Since $D$ is self-adjoint, $D^0$ and $D^1$ are adjoints of each other and the index of $D^0$ is written as
\begin{equation}
\text{ind}(D^0)=\text{dim}(\text{ker } D^0)-\text{dim}(\text{ker } D^1)
\end{equation}
where $\text{ker }D^0$ denotes the kernel of $D^0$.

For $Cl_k$-Dirac bundles, we denote the Dirac operator as $\displaystyle{\not}D_k$ and it commutes with the $Cl_k$-action which means that it is a $Cl_k$-linear operator. If the bundle is $\mathbb{Z}_2$-graded, then the $Cl_k$-Dirac operator is
\begin{equation}
\displaystyle{\not}D_k=\left(\begin{array}{cc}
0 & \displaystyle{\not}D^1_k \\
\displaystyle{\not}D^0_k & 0 \\
\end{array}\right)
\end{equation}
The index of the $Cl_k$-Dirac operator is given by
\begin{equation}
\text{ind}\displaystyle{\not}D^0_k=\text{dim}(\text{ker }\displaystyle{\not}D^0_k)-\text{dim}(\text{ker }\displaystyle{\not}D^1_k)
\end{equation}
The constructions for $Cl^*_k$-Dirac bundles are similar. By defining $Cl^*_k$-Dirac operators on $Cl^*_k$-Dirac bundles, we can construct the table of $Cl^*_{s-n}$-Dirac operators from the table of $Cl^*_{s-n}$-bundles defined in subsection 4.A as follows

\quad\\
{\centering{
\begin{tabular}{c c c c c c c c c}

% after \\: \hline or \cline{col1-col2} ...
\hline \hline
$\displaystyle{\not}D_{s-n (\text{mod }8)}$ & $n=0$ & $1$ & $2$ & $3$ & $4$ & $5$ & $6$ & $7$ \\ \hline
$s=0$ & $\displaystyle{\not}D_0$ & $\displaystyle{\not}D_7$ & $\displaystyle{\not}D_6$ & $\displaystyle{\not}D_5$ & $\displaystyle{\not}D_4$ & $\displaystyle{\not}D_3$ & $\displaystyle{\not}D_2$ & $\displaystyle{\not}D_1$ \\
$1$ & $\displaystyle{\not}D_1$ & $\displaystyle{\not}D_0$ & $\displaystyle{\not}D_7$ & $\displaystyle{\not}D_6$ & $\displaystyle{\not}D_5$ & $\displaystyle{\not}D_4$ & $\displaystyle{\not}D_3$ & $\displaystyle{\not}D_2$ \\
$2$ & $\displaystyle{\not}D_2$ & $\displaystyle{\not}D_1$ & $\displaystyle{\not}D_0$ & $\displaystyle{\not}D_7$ & $\displaystyle{\not}D_6$ & $\displaystyle{\not}D_5$ & $\displaystyle{\not}D_4$ & $\displaystyle{\not}D_3$ \\
$3$ & $\displaystyle{\not}D_3$ & $\displaystyle{\not}D_2$ & $\displaystyle{\not}D_1$ & $\displaystyle{\not}D_0$ & $\displaystyle{\not}D_7$ & $\displaystyle{\not}D_6$ & $\displaystyle{\not}D_5$ & $\displaystyle{\not}D_4$ \\
$4$ & $\displaystyle{\not}D_4$ & $\displaystyle{\not}D_3$ & $\displaystyle{\not}D_2$ & $\displaystyle{\not}D_1$ & $\displaystyle{\not}D_0$ & $\displaystyle{\not}D_7$ & $\displaystyle{\not}D_6$ & $\displaystyle{\not}D_5$ \\
$5$ & $\displaystyle{\not}D_5$ & $\displaystyle{\not}D_4$ & $\displaystyle{\not}D_3$ & $\displaystyle{\not}D_2$ & $\displaystyle{\not}D_1$ & $\displaystyle{\not}D_0$ & $\displaystyle{\not}D_7$ & $\displaystyle{\not}D_6$ \\
$6$ & $\displaystyle{\not}D_6$ & $\displaystyle{\not}D_5$ & $\displaystyle{\not}D_4$ & $\displaystyle{\not}D_3$ & $\displaystyle{\not}D_2$ & $\displaystyle{\not}D_1$ & $\displaystyle{\not}D_0$ & $\displaystyle{\not}D_7$ \\
$7$ & $\displaystyle{\not}D_7$ & $\displaystyle{\not}D_6$ & $\displaystyle{\not}D_5$ & $\displaystyle{\not}D_4$ & $\displaystyle{\not}D_3$ & $\displaystyle{\not}D_2$ & $\displaystyle{\not}D_1$ & $\displaystyle{\not}D_0$ \\ \hline \hline

\end{tabular}}
\quad\\
\quad\\}

\subsection{Index theorem for $Cl_k$-Dirac operators}

If the Dirac bundle $S$ is the Clifford bundle on $M$, then the Dirac operator $D$ corresponds to the Hodge-de Rham operator and its square $D^2=\Delta$ is the Hodge Laplacian. The differential forms that are in the kernel of the Hodge Laplacian are called harmonic forms. Similarly, if $S$ is the spinor bundle on $M$, then $D$ is the usual Dirac operator on spinors and the spinors that are in the kernel of $D$, namely the spinors that satisfy $D\psi=0$, are called harmonic spinors since $\text{ker }D=\text{ker }D^2$ for any compact Riemannian manifold $M$. These definitions can be extended to the case of $Cl_k$-Dirac operators and we denote the kernel of $\displaystyle{\not}D_k$, that is the harmonic space, as $\textbf{H}_k=\text{ker }\displaystyle{\not}D_k$.

Now, we consider the index theorem for $Cl_k$-Dirac operators. The analytic index of $\displaystyle{\not}D_k$ defined in (21) can be written in terms of the dimension of the harmonic space $\textbf{H}_k$ and characteristic classes of the bundle \cite{Lawson Michelsohn}. The Atiyah-Singer index theorem for $\displaystyle{\not}D_k$ gives the following equality
\begin{eqnarray}
\text{ind}(\displaystyle{\not}D_k)&=&\left\{
                                                                               \begin{array}{ll}
                                                                                 \text{dim}_{\mathbb{C}}\textbf{H}_k (\text{mod }2), & \hbox{ for $k\equiv 1$  $(\text{mod }8)$} \\
                                                                                 \text{dim}_{\mathbb{H}}\textbf{H}_k (\text{mod }2), & \hbox{ for $k\equiv 2$  $(\text{mod }8)$} \\
                                                                                 \frac{1}{2}\widehat{A}(M), & \hbox{ for $k\equiv 4$  $(\text{mod }8)$} \\
                                                                                 \widehat{A}(M), & \hbox{ for $k\equiv 0$  $(\text{mod }8)$}
                                                                               \end{array}
                                                                             \right.
\end{eqnarray}
where $\text{dim}_{\mathbb{C}}$ and $\text{dim}_{\mathbb{H}}$ denote the complex and quaternionic dimensions, respectively. $\widehat{A}(M)$ is the $\widehat{A}$-genus of the manifold $M$ and it is defined as a power series expansion
\begin{equation}
\widehat{A}(M)=\prod_{i=1}^n \frac{x_i/2}{\text{sinh}(x_i/2)}.
\end{equation}
It can be written in terms of the Pontrjagin classes $p_i$ as follows
\begin{eqnarray}
\widehat{A}(M)&=&1-\frac{1}{24}p_1+\frac{1}{5760}\left(7p_1^2-4p_2\right)+\frac{1}{967680}\left(-31p_1^3+44p_1p_2-16p_3\right)+...
\end{eqnarray}
$\widehat{A}$-genus is an integer number for compact manifolds and it is an even integer for the dimensions $4 (\text{mod }8)$. So, the index of $\displaystyle{\not}D_k$ takes values in $\mathbb{Z}_2$ for $k\equiv 1 \text{ and }2 (\text{mod }8)$ and in $\mathbb{Z}$ for $k\equiv 0 \text{ and }4 (\text{mod}8)$.

Then, we can write the table of $Cl^*_{s-n}$-Dirac operators in subsection 4.B in terms of the index of $Cl^*_{s-n}$-Dirac operators in the following way

\quad\\
{\centering{
\begin{tabular}{c c c c c c c c c}

% after \\: \hline or \cline{col1-col2} ...
\hline \hline
$\text{ind}(\displaystyle{\not}D_{s-n (\text{mod }8)})$ & $n=0$ & $1$ & $2$ & $3$ & $4$ & $5$ & $6$ & $7$ \\ \hline
$s=0$ & $\widehat{A}(M)$ & $0$ & $0$ & $0$ & $\frac{1}{2}\widehat{A}(M)$ & $0$ & $\text{dim}_{\mathbb{H}}\textbf{H}_k$ & $\text{dim}_{\mathbb{C}}\textbf{H}_k$ \\
$1$ & $\text{dim}_{\mathbb{C}}\textbf{H}_k$ & $\widehat{A}(M)$ & $0$ & $0$ & $0$ & $\frac{1}{2}\widehat{A}(M)$ & $0$ & $\text{dim}_{\mathbb{H}}\textbf{H}_k$ \\
$2$ & $\text{dim}_{\mathbb{H}}\textbf{H}_k$ & $\text{dim}_{\mathbb{C}}\textbf{H}_k$ & $\widehat{A}(M)$ & $0$ & $0$ & $0$ & $\frac{1}{2}\widehat{A}(M)$ & $0$ \\
$3$ & $0$ & $\text{dim}_{\mathbb{H}}\textbf{H}_k$ & $\text{dim}_{\mathbb{C}}\textbf{H}_k$ & $\widehat{A}(M)$ & $0$ & $0$ & $0$ & $\frac{1}{2}\widehat{A}(M)$ \\
$4$ & $\frac{1}{2}\widehat{A}(M)$ & $0$ & $\text{dim}_{\mathbb{H}}\textbf{H}_k$ & $\text{dim}_{\mathbb{C}}\textbf{H}_k$ & $\widehat{A}(M)$ & $0$ & $0$ & $0$ \\
$5$ & $0$ & $\frac{1}{2}\widehat{A}(M)$ & $0$ & $\text{dim}_{\mathbb{H}}\textbf{H}_k$ & $\text{dim}_{\mathbb{C}}\textbf{H}_k$ & $\widehat{A}(M)$ & $0$ & $0$ \\
$6$ & $0$ & $0$ & $\frac{1}{2}\widehat{A}(M)$ & $0$ & $\text{dim}_{\mathbb{H}}\textbf{H}_k$ & $\text{dim}_{\mathbb{C}}\textbf{H}_k$ & $\widehat{A}(M)$ & $0$ \\
$7$ & $0$ & $0$ & $0$ & $\frac{1}{2}\widehat{A}(M)$ & $0$ & $\text{dim}_{\mathbb{H}}\textbf{H}_k$ & $\text{dim}_{\mathbb{C}}\textbf{H}_k$ & $\widehat{A}(M)$ \\ \hline \hline

\end{tabular}}
\quad\\
\quad\\
\quad\\}

The Atiyah-Singer index theorem in (22) attaches different topological invariants to $Cl^*_k$-Dirac operators $\displaystyle{\not}D_k$ for different values of $k$. This means that we have non-trivial topological classes for different values of $s$ and $n$ which have non-zero index. For the cases of zero index, there are only trivial classes. In that way, we obtain a table of topological equivalence classes of $Cl^*_k$-Dirac bundles from the Clifford chessboard by using the index theorems.

\subsection{Relation with $KO$-groups}

The Dirac operator $D$ on a real Dirac bundle is a self-adjoint operator with finite dimensional kernel and cokernel. So, it is an element of the set of real Fredholm operators $\mathfrak{F}_{\mathbb{R}}$. The index of Fredholm operators is constant on connected components of $\mathfrak{F}_{\mathbb{R}}$ and we have the following isomorphism
\begin{equation}
\text{ind}:\pi_0(\mathfrak{F}_{\mathbb{R}})\stackrel{\simeq}{\longrightarrow}\mathbb{Z}
\end{equation}
Moreover, $\mathfrak{F}_{\mathbb{R}}$ is a classifying space for $KO$-groups. The stable equivalence classes of real vector bundles on a manifold $M$ are characterized by $KO$-groups and they are real equivalences of the $K$-groups of complex vector bundles. For a Haussdorf space $X$, index map defines an isomorphism \cite{Lawson Michelsohn}
\begin{eqnarray}
\text{ind}:[X,\mathfrak{F}_{\mathbb{R}}]\longrightarrow KO(X)
\end{eqnarray}
where $[X,\mathfrak{F}_{\mathbb{R}}]$ denotes the homotopy classes of maps between $X$ and $\mathfrak{F}_{\mathbb{R}}$ and $KO(X)$ is the $KO$-group on $X$ that classifies the stably equivalent real vector bundles on $X$. So, this gives an isomorphism between the homotopy groups of $\mathfrak{F}_{\mathbb{R}}$ and the $KO$-groups at the point
\begin{equation}
\text{ind}:\pi_k(\mathfrak{F}_{\mathbb{R}})\stackrel{\simeq}{\longrightarrow}KO^{-k}(\text{pt})\equiv \widetilde{KO}(S^k)
\end{equation}
where $KO^{-k}(X)$ is equivalent to the reduced $KO$-group of $k$-fold suspension $\Sigma^kX$ of $X$; $KO^{-k}(X)\equiv \widetilde{KO}(\Sigma^kX)$. The reduced $KO$-group $\widetilde{KO}(X)$ is defined as the kernel of the map $KO(X)\longrightarrow KO(\text{pt})\cong \mathbb{Z}$ and $KO(\text{pt})$ is the $KO$-group at the point. Since the $k$-fold suspension of the point is equivalent to the $k$-sphere $\Sigma^k(\text{pt})\equiv S^k$, we have $KO^{-k}(\text{pt})\equiv \widetilde{KO}(S^k)$.

Let us denote the subset of Fredholm operators which are $\mathbb{Z}_2$-graded, $Cl_k$-linear and self-adjoint as $\mathfrak{F}_k$. Then, the index of the operators in $\mathfrak{F}_k$ is constant on connected components of $\mathfrak{F}_k$
\[
\text{ind}:\pi_0(\mathfrak{F}_k)\longrightarrow KO^{-k}(\text{pt})
\]
So, $\mathfrak{F}_k$ is a classifying space for $KO^{-k}$ and for a Haussdorf space $X$ and for any $k$, there is an isomorphism
\[
\text{ind}:[X,\mathfrak{F}_k]\longrightarrow KO^{-k}(X)
\]
This means that the index of the $Cl_k$-Dirac operator $\displaystyle{\not}D_k$ takes values in the group $KO^{-k}(\text{pt})$ and we can write the table of the index of $Cl^*_{s-n}$-Dirac operators in subsection 4.C as the table of $KO^{-(s-n)}(\text{pt})$ groups

\quad\\
{\centering{
\begin{tabular}{c c c c c c c c c}

% after \\: \hline or \cline{col1-col2} ...
\hline \hline
$KO^{-(s-n) (\text{mod }8)}(\text{pt})$ & $n=0$ & $1$ & $2$ & $3$ & $4$ & $5$ & $6$ & $7$ \\ \hline
$s=0$ & $KO(\text{pt})$ & $KO^{-7}(\text{pt})$ & $KO^{-6}(\text{pt})$ & $KO^{-5}(\text{pt})$ & $KO^{-4}(\text{pt})$ & $KO^{-3}(\text{pt})$ & $KO^{-2}(\text{pt})$ & $KO^{-1}(\text{pt})$ \\
$1$ & $KO^{-1}(\text{pt})$ & $KO(\text{pt})$ & $KO^{-7}(\text{pt})$ & $KO^{-6}(\text{pt})$ & $KO^{-5}(\text{pt})$ & $KO^{-4}(\text{pt})$ & $KO^{-3}(\text{pt})$ & $KO^{-2}(\text{pt})$ \\
$2$ & $KO^{-2}(\text{pt})$ & $KO^{-1}(\text{pt})$ & $KO(\text{pt})$ & $KO^{-7}(\text{pt})$ & $KO^{-6}(\text{pt})$ & $KO^{-5}(\text{pt})$ & $KO^{-4}(\text{pt})$ & $KO^{-3}(\text{pt})$ \\
$3$ & $KO^{-3}(\text{pt})$ & $KO^{-2}(\text{pt})$ & $KO^{-1}(\text{pt})$ & $KO(\text{pt})$ & $KO^{-7}(\text{pt})$ & $KO^{-6}(\text{pt})$ & $KO^{-5}(\text{pt})$ & $KO^{-4}(\text{pt})$ \\
$4$ & $KO^{-4}(\text{pt})$ & $KO^{-3}(\text{pt})$ & $KO^{-2}(\text{pt})$ & $KO^{-1}(\text{pt})$ & $KO(\text{pt})$ & $KO^{-7}(\text{pt})$ & $KO^{-6}(\text{pt})$ & $KO^{-5}(\text{pt})$ \\
$5$ & $KO^{-5}(\text{pt})$ & $KO^{-4}(\text{pt})$ & $KO^{-3}(\text{pt})$ & $KO^{-2}(\text{pt})$ & $KO^{-1}(\text{pt})$ & $KO(\text{pt})$ & $KO^{-7}(\text{pt})$ & $KO^{-6}(\text{pt})$ \\
$6$ & $KO^{-6}(\text{pt})$ & $KO^{-5}(\text{pt})$ & $KO^{-4}(\text{pt})$ & $KO^{-3}(\text{pt})$ & $KO^{-2}(\text{pt})$ & $KO^{-1}(\text{pt})$ & $KO(\text{pt})$ & $KO^{-7}(\text{pt})$ \\
$7$ & $KO^{-7}(\text{pt})$ & $KO^{-6}(\text{pt})$ & $KO^{-5}(\text{pt})$ & $KO^{-4}(\text{pt})$ & $KO^{-3}(\text{pt})$ & $KO^{-2}(\text{pt})$ & $KO^{-1}(\text{pt})$ & $KO(\text{pt})$ \\ \hline \hline

\end{tabular}}
\quad\\
\quad\\
\quad}

Since the $KO$-groups of the point is given as in the following table

\quad\\
{\centering{
\begin{tabular}{c c c c c c c c c}

% after \\: \hline or \cline{col1-col2} ...
\hline \hline
$k$ & $0$ & $1$ & $2$ & $3$ & $4$ & $5$ & $6$ & $7$ \\ \hline
$KO^{-k}(\text{pt})$ & $\mathbb{Z}$\, & $\mathbb{Z}_2$ & $\mathbb{Z}_2$ & $0\,\,$ & $\mathbb{Z}$\,\, & $0\,\,$ & $0\,\,$ & $0$ \\ \hline \hline

\end{tabular}}
\quad\\
\quad\\
\quad\\}
we obtain the table of $KO^{-(s-n)}(\text{pt})$ groups as

\quad\\
{\centering{
\begin{tabular}{c c c c c c c c c}

% after \\: \hline or \cline{col1-col2} ...
\hline \hline
$KO^{-(s-n) (\text{mod }8)}(\text{pt})$ & $n=0$ & $1$ & $2$ & $3$ & $4$ & $5$ & $6$ & $7$ \\ \hline
$s=0$ & $\mathbb{Z}$ & $0$ & $0$ & $0$ & $\mathbb{Z}$ & $0$ & $\mathbb{Z}_2$ & $\mathbb{Z}_2$ \\
$1$ & $\mathbb{Z}_2$ & $\mathbb{Z}$ & $0$ & $0$ & $0$ & $\mathbb{Z}$ & $0$ & $\mathbb{Z}_2$ \\
$2$ & $\mathbb{Z}_2$ & $\mathbb{Z}_2$ & $\mathbb{Z}$ & $0$ & $0$ & $0$ & $\mathbb{Z}$ & $0$ \\
$3$ & $0$ & $\mathbb{Z}_2$ & $\mathbb{Z}_2$ & $\mathbb{Z}$ & $0$ & $0$ & $0$ & $\mathbb{Z}$ \\
$4$ & $\mathbb{Z}$ & $0$ & $\mathbb{Z}_2$ & $\mathbb{Z}_2$ & $\mathbb{Z}$ & $0$ & $0$ & $0$ \\
$5$ & $0$ & $\mathbb{Z}$ & $0$ & $\mathbb{Z}_2$ & $\mathbb{Z}_2$ & $\mathbb{Z}$ & $0$ & $0$ \\
$6$ & $0$ & $0$ & $\mathbb{Z}$ & $0$ & $\mathbb{Z}_2$ & $\mathbb{Z}_2$ & $\mathbb{Z}$ & $0$ \\
$7$ & $0$ & $0$ & $0$ & $\mathbb{Z}$ & $0$ & $\mathbb{Z}_2$ & $\mathbb{Z}_2$ & $\mathbb{Z}$ \\ \hline \hline

\end{tabular}}
\quad\\
\quad\\
\quad}

and this is compatible with the fact that the index of $\displaystyle{\not}D_k$ takes values in $\mathbb{Z}$ and $\mathbb{Z}_2$ as stated in subsection 4.C.

\subsection{Isomorphism between $KO$-groups and Grothendieck groups}

Now, we consider the Atiyah-Bott-Shapiro (ABS) isomorphism which relates $KO$-groups with the real representations of Clifford algebras \cite{Atiyah Bott Shapiro}. Let us denote the group of equivalence classes of irreducible real representations of $Cl_k$ as $\mathfrak{M}_k$ which is called the Grothendieck group. This is the free abelian group generated by the distinct irreducible representations over $\mathbb{R}$. For different values of $k$, we have the Grothendieck groups as in the following table \cite{Lawson Michelsohn}

\quad\\
{\centering{
\begin{tabular}{c c c c c c c c c}

% after \\: \hline or \cline{col1-col2} ...
\hline \hline
$k$ & $0$ & $1$ & $2$ & $3$ & $4$ & $5$ & $6$ & $7$ \\ \hline
$Cl_k$ & \,\,\,\,$\mathbb{R}$\,\,\,\, & \,\,\,\,$\mathbb{C}$\,\,\,\, & \,\,\,\,$\mathbb{H}$\,\,\,\, & \,\,\,\,$\mathbb{H}\oplus\mathbb{H}$\,\,\,\, & \,\,\,\,$\mathbb{H}(2)$\,\,\,\, & \,\,\,\,$\mathbb{C}(4)$\,\,\,\, & \,\,\,\,$\mathbb{R}(8)$\,\,\,\, & \,\,\,\,$\mathbb{R}(8)\oplus\mathbb{R}(8)$\,\,\,\, \\
$\mathfrak{M}_k$ & \,\,\,\,$\mathbb{Z}$\,\,\,\, & \,\,\,\,$\mathbb{Z}$\,\,\,\, & \,\,\,\,$\mathbb{Z}$\,\,\,\, & \,\,\,\,$\mathbb{Z}\oplus\mathbb{Z}$\,\,\,\, & \,\,\,\,$\mathbb{Z}$\,\,\,\, & \,\,\,\,$\mathbb{Z}$\,\,\,\, & \,\,\,\,$\mathbb{Z}\,\,\,\,$ & \,\,\,\,$\mathbb{Z}\oplus\mathbb{Z}$\,\,\,\, \\ \hline \hline

\end{tabular}}
\quad\\
\quad\\
\quad\\}

The inclusion $i:\mathbb{R}^k\hookrightarrow\mathbb{R}^{k+1}$ given by $i(x_1,...,x_k)=(x_1,...,x_k,0)$ induces an algebra homomorphism $i_*:Cl_k\longrightarrow Cl_{k+1}$. By restricting the action from $Cl_{k+1}$ to $Cl_k$, we obtain the homomorphism $i^*:\mathfrak{M}_{k+1}\longrightarrow\mathfrak{M}_k$. So, we can define the groups $\mathfrak{M}_k/i^*\mathfrak{M}_{k+1}$. In this quotient group, a representation that can be obtained by restricting a representation of $Cl_{k+1}$ to $Cl_k$ is equivalent to zero. Similarly, the Grothendieck group of real $\mathbb{Z}_2$-graded modules over $Cl_k$ is denoted by $\widehat{\mathfrak{M}}_k$ and there is the natural isomorphism $\widehat{\mathfrak{M}}_k=\mathfrak{M}_{k-1}$ that comes from (13). This implies the following isomorphism
\[
\widehat{\mathfrak{M}}_k/i^*\widehat{\mathfrak{M}}_{k+1}\cong\mathfrak{M}_{k-1}/i^*\mathfrak{M}_k
\]
From the groups $\mathfrak{M}_k$ defined in the above table, we obtain
\begin{eqnarray}
\widehat{\mathfrak{M}}_k/i^*\widehat{\mathfrak{M}}_{k+1}&\cong&\left\{
                                                                               \begin{array}{ll}
                                                                                 \mathbb{Z}, & \hbox{ for $k\equiv 0$ \text{or}  $4(\text{mod }8)$} \\
                                                                                 \mathbb{Z}_2, & \hbox{ for $k\equiv 1$ \text{or}  $2(\text{mod }8)$} \\
                                                                                 0, & \hbox{ otherwise}
                                                                               \end{array}
                                                                             \right.
\end{eqnarray}
where the $\mathbb{Z}$ groups arise when $\mathfrak{M}_{k-1}$ correspond to $\mathbb{Z}\oplus\mathbb{Z}$ and the $\mathbb{Z}_2$ groups arise when the dimension of the representation of $Cl_k$ is double of the dimension of the representation of $Cl_{k-1}$ \cite{Stone Chiu Roy}. The graded tensor product of modules gives a multiplication in
$\widehat{\mathfrak{M}}_k/i^*\widehat{\mathfrak{M}}_{k+1}$ and this
defines the graded ring
$\widehat{\mathfrak{M}}_*/i^*\widehat{\mathfrak{M}}_{*+1}\equiv
\bigoplus_{k\geq
0}\widehat{\mathfrak{M}}_k/i^*\widehat{\mathfrak{M}}_{k+1}$.

ABS isomorphism defines an isomorphism between the $KO$-groups at the
point and the quotients of Grothendieck groups in the following way
\[
\phi:\widehat{\mathfrak{M}}_*/i^*\widehat{\mathfrak{M}}_{*+1}\stackrel{\simeq}{\longrightarrow}KO^{-*}(\text{pt})
\]
This can easily be seen from the definitions in this and previous subsections. So, we can write the table of $KO^{-(s-n)}(\text{pt})$ groups in subsection 4.D as the table of $\widehat{\mathfrak{M}}_{s-n}/i^*\widehat{\mathfrak{M}}_{s-n+1}$ groups as follows

\quad\\
{\centering{
\begin{tabular}{c c c c c c c c c}

% after \\: \hline or \cline{col1-col2} ...
\hline \hline
$\widehat{\mathfrak{M}}_{s-n (\text{mod }8)}/i^*\widehat{\mathfrak{M}}_{s-n+1(\text{mod }8)}$ & $n=0$ & $1$ & $2$ & $3$ & $4$ & $5$ & $6$ & $7$ \\ \hline
$s=0$ & $\mathbb{Z}$ & $0$ & $0$ & $0$ & $\mathbb{Z}$ & $0$ & $\mathbb{Z}_2$ & $\mathbb{Z}_2$ \\
$1$ & $\mathbb{Z}_2$ & $\mathbb{Z}$ & $0$ & $0$ & $0$ & $\mathbb{Z}$ & $0$ & $\mathbb{Z}_2$ \\
$2$ & $\mathbb{Z}_2$ & $\mathbb{Z}_2$ & $\mathbb{Z}$ & $0$ & $0$ & $0$ & $\mathbb{Z}$ & $0$ \\
$3$ & $0$ & $\mathbb{Z}_2$ & $\mathbb{Z}_2$ & $\mathbb{Z}$ & $0$ & $0$ & $0$ & $\mathbb{Z}$ \\
$4$ & $\mathbb{Z}$ & $0$ & $\mathbb{Z}_2$ & $\mathbb{Z}_2$ & $\mathbb{Z}$ & $0$ & $0$ & $0$ \\
$5$ & $0$ & $\mathbb{Z}$ & $0$ & $\mathbb{Z}_2$ & $\mathbb{Z}_2$ & $\mathbb{Z}$ & $0$ & $0$ \\
$6$ & $0$ & $0$ & $\mathbb{Z}$ & $0$ & $\mathbb{Z}_2$ & $\mathbb{Z}_2$ & $\mathbb{Z}$ & $0$ \\
$7$ & $0$ & $0$ & $0$ & $\mathbb{Z}$ & $0$ & $\mathbb{Z}_2$ & $\mathbb{Z}_2$ & $\mathbb{Z}$ \\ \hline \hline

\end{tabular}}
\quad\\
\quad\\
\quad}

and the Clifford chessboard in section 3 turns into the table of quotients of Grothendieck groups.

\subsection{From Grothendieck groups to symmetric spaces}

$KO$-groups and Grothendieck groups are in relation with Cartan symmetric spaces through their connected components. If we consider the following sequence of groups
\[
O(16n)\longrightarrow U(8n)\longrightarrow Sp(4n)\longrightarrow Sp(2n)\times Sp(2n)\longrightarrow Sp(2n)\longrightarrow U(2n)\longrightarrow O(2n)\longrightarrow O(n)\times O(n)\longrightarrow O(n)
\]
and denote every element in the sequence by $G_i$, then the coset spaces $R_i\equiv G_i/G_{i+1}$ correspond to the eight of ten symmetric spaces of Cartan. These symmetric spaces and the groups that give their connected components are given as follows \cite{Stone Chiu Roy}

\quad\\
{\centering{
\begin{tabular}{c c c}

% after \\: \hline or \cline{col1-col2}
\hline \hline
$R_i$ & $\text{symmetric spaces}$ & $\pi_0(R_i)$ \\ \hline
$R_0$ & $O(2n)/O(n)\times O(n)$ & $\mathbb{Z}$ \\
$R_1$ & $O(n)\times O(n)/O(n)$ & $\mathbb{Z}_2$ \\
$R_2$ & $O(2n)/U(n)$ & $\mathbb{Z}_2$ \\
$R_3$ & $U(2n)/Sp(n)$ & $0$ \\
$R_4$ & $Sp(2n)/Sp(n)\times Sp(n)$ & $\mathbb{Z}$ \\
$R_5$ & $Sp(n)\times Sp(n)/Sp(n)$ & $0$ \\
$R_6$ & $Sp(n)/U(n)$ & $0$ \\
$R_7$ & $U(n)/O(n)$ & $0$ \\ \hline \hline

\end{tabular}}
\quad\\
\quad\\
\quad\\}

As can be seen from the last column of the table, the connected components of the symmetric spaces are in one-to-one correspondence with $KO$-groups at the point and hence to the Grothendieck groups. Indeed, Cartan symmetric spaces correspond to the classifying spaces of vector bundles and we have the following isomorphism for a Haussdorf space $X$
\[
KO^{-k}(X)\cong [X,R_k]
\]
and so
\[
KO^{-k}(\text{pt})\cong \pi_0(R_k).
\]
From the ABS isomorphism in subsection 4.E, we also have
\[
\widehat{\mathfrak{M}}_k/i^*\widehat{\mathfrak{M}}_{k+1}\cong \pi_0(R_k).
\]
Hence, we can write the table of the quotients of Grothendieck groups in subsection 4.E in terms of the connected components of the symmetric spaces as follows

\quad\\
{\centering{
\begin{tabular}{c c c c c c c c c}

% after \\: \hline or \cline{col1-col2} ...
\hline \hline
$\pi_0(R_{s-n(\text{mod }8)})$ & $n=0$ & $1$ & $2$ & $3$ & $4$ & $5$ & $6$ & $7$ \\ \hline
$s=0$ & $\pi_0(R_0)$ & $\pi_0(R_7)$ & $\pi_0(R_6)$ & $\pi_0(R_5)$ & $\pi_0(R_4)$ & $\pi_0(R_3)$ & $\pi_0(R_2)$ & $\pi_0(R_1)$ \\
$1$ & $\pi_0(R_1)$ & $\pi_0(R_0)$ & $\pi_0(R_7)$ & $\pi_0(R_6)$ & $\pi_0(R_5)$ & $\pi_0(R_4)$ & $\pi_0(R_3)$ & $\pi_0(R_2)$ \\
$2$ & $\pi_0(R_2)$ & $\pi_0(R_1)$ & $\pi_0(R_0)$ & $\pi_0(R_7)$ & $\pi_0(R_6)$ & $\pi_0(R_5)$ & $\pi_0(R_4)$ & $\pi_0(R_3)$ \\
$3$ & $\pi_0(R_3)$ & $\pi_0(R_2)$ & $\pi_0(R_1)$ & $\pi_0(R_0)$ & $\pi_0(R_7)$ & $\pi_0(R_6)$ & $\pi_0(R_5)$ & $\pi_0(R_4)$ \\
$4$ & $\pi_0(R_4)$ & $\pi_0(R_3)$ & $\pi_0(R_2)$ & $\pi_0(R_1)$ & $\pi_0(R_0)$ & $\pi_0(R_7)$ & $\pi_0(R_6)$ & $\pi_0(R_5)$ \\
$5$ & $\pi_0(R_5)$ & $\pi_0(R_4)$ & $\pi_0(R_3)$ & $\pi_0(R_2)$ & $\pi_0(R_1)$ & $\pi_0(R_0)$ & $\pi_0(R_7)$ & $\pi_0(R_6)$ \\
$6$ & $\pi_0(R_6)$ & $\pi_0(R_5)$ & $\pi_0(R_4)$ & $\pi_0(R_3)$ & $\pi_0(R_2)$ & $\pi_0(R_1)$ & $\pi_0(R_0)$ & $\pi_0(R_7)$ \\
$7$ & $\pi_0(R_7)$ & $\pi_0(R_6)$ & $\pi_0(R_5)$ & $\pi_0(R_4)$ & $\pi_0(R_3)$ & $\pi_0(R_2)$ & $\pi_0(R_1)$ & $\pi_0(R_0)$ \\ \hline \hline

\end{tabular}}
\quad\\
\quad\\
\quad}

However, the homotopy groups of symmetric spaces have the property $\pi_n(R_k)=\pi_{n+1}(R_{k-1})$, so we have
\[
\pi_0(R_{s-n(\text{mod }8)})=\pi_{8-n(\text{mod }8)}(R_s)
\]
and the table turns into

\quad\\
{\centering{
\begin{tabular}{c c c c c c c c c}

% after \\: \hline or \cline{col1-col2} ...
\hline \hline
$\pi_{8-n(\text{mod }8)}(R_s)$ & $n=0$ & $1$ & $2$ & $3$ & $4$ & $5$ & $6$ & $7$ \\ \hline
$s=0$ & $\mathbb{Z}$ & $0$ & $0$ & $0$ & $\mathbb{Z}$ & $0$ & $\mathbb{Z}_2$ & $\mathbb{Z}_2$ \\
$1$ & $\mathbb{Z}_2$ & $\mathbb{Z}$ & $0$ & $0$ & $0$ & $\mathbb{Z}$ & $0$ & $\mathbb{Z}_2$ \\
$2$ & $\mathbb{Z}_2$ & $\mathbb{Z}_2$ & $\mathbb{Z}$ & $0$ & $0$ & $0$ & $\mathbb{Z}$ & $0$ \\
$3$ & $0$ & $\mathbb{Z}_2$ & $\mathbb{Z}_2$ & $\mathbb{Z}$ & $0$ & $0$ & $0$ & $\mathbb{Z}$ \\
$4$ & $\mathbb{Z}$ & $0$ & $\mathbb{Z}_2$ & $\mathbb{Z}_2$ & $\mathbb{Z}$ & $0$ & $0$ & $0$ \\
$5$ & $0$ & $\mathbb{Z}$ & $0$ & $\mathbb{Z}_2$ & $\mathbb{Z}_2$ & $\mathbb{Z}$ & $0$ & $0$ \\
$6$ & $0$ & $0$ & $\mathbb{Z}$ & $0$ & $\mathbb{Z}_2$ & $\mathbb{Z}_2$ & $\mathbb{Z}$ & $0$ \\
$7$ & $0$ & $0$ & $0$ & $\mathbb{Z}$ & $0$ & $\mathbb{Z}_2$ & $\mathbb{Z}_2$ & $\mathbb{Z}$ \\ \hline \hline

\end{tabular}}
\quad\\
\quad\\
\quad}

This shows that every row in the table denoted by $s$ corresponds to a Cartan symmetric space $R_s$ and every column denoted by $n$ corresponds to the $(8-n)$th homotopy group of the symmetric space.

\subsection{From symmetric spaces to periodic table}

Eight symmetric spaces defined in the previous subsection are in one-to-one correspondence with the extensions of some real Clifford algebras. By starting with a real Clifford algebra and answering the question how many different types of generators can be added to the existing ones gives the symmetric spaces \cite{Morimoto Furusaki}. The symmetric spaces and corresponding Clifford algebra extensions are given as follows

\quad\\
{\centering{
\begin{tabular}{c c c}

% after \\: \hline or \cline{col1-col2}
\hline \hline
$R_i$ & $$ & $\text{Extension}$ \\ \hline
$R_0$ & $$ & $Cl_{0,2}\longrightarrow Cl_{1,2}$ \\
$R_1$ & $$ & $Cl_{1,2}\longrightarrow Cl_{1,3}$ \\
$R_2$ & $$ & $Cl_{0,2}\longrightarrow Cl_{0,3}$ \\
$R_3$ & $$ & $Cl_{0,3}\longrightarrow Cl_{0,4}$ \\
$R_4$ & $$ & $Cl_{2,0}\longrightarrow Cl_{3,0}$ \\
$R_5$ & $$ & $Cl_{3,0}\longrightarrow Cl_{3,1}$ \\
$R_6$ & $$ & $Cl_{2,0}\longrightarrow Cl_{2,1}$ \\
$R_7$ & $$ & $Cl_{2,1}\longrightarrow Cl_{2,2}$ \\ \hline \hline

\end{tabular}}
\quad\\
\quad\\
\quad\\}

These Clifford algebras and their generators can be represented by some symmetry transformations of fermionic Hamiltonians. They are constructed in terms of the time-reversal operator $T$ and the charge-conjugation operator $C$ of free fermion Hamiltonians. The extension problem of Clifford algebras can be described by the addition of the free-fermion Hamiltonian $H$ to the existing Clifford algebra generators constructed out of $T$ and $C$ \cite{Morimoto Furusaki}. By defining $S=TC$ as the chiral symmetry operator and considering the fact that $T^2$ and $C^2$ can have values $\pm1$, the table of symmetric spaces turns into the Altland-Zirnbauer symmetry classes of free fermion Hamiltonians \cite{Altland Zirnbauer}. So, the table of homotopy groups of symmetric spaces in subsection 4.F gives the periodic table of topological insulators and superconductors for real classes

\quad\\
{\centering{
\begin{tabular}{c | c c c | c c c c c c c c}

% after \\: \hline or \cline{col1-col2} ...
\hline \hline
$\text{label}$ & $T$ & $C$ & $S$ & $0$ & $1$ & $2$ & $3$ & $4$ & $5$ & $6$ & $7$ \\ \hline
$\text{AI}$ & $1$ & $0$ & $0$ & $\mathbb{Z}$ & $0$ & $0$ & $0$ & $\mathbb{Z}$ & $0$ & $\mathbb{Z}_2$ & $\mathbb{Z}_2$ \\
$\text{BDI}$ & $1$ & $1$ & $1$ & $\mathbb{Z}_2$ & $\mathbb{Z}$ & $0$ & $0$ & $0$ & $\mathbb{Z}$ & $0$ & $\mathbb{Z}_2$ \\
$\text{D}$ & $0$ & $1$ & $0$ & $\mathbb{Z}_2$ & $\mathbb{Z}_2$ & $\mathbb{Z}$ & $0$ & $0$ & $0$ & $\mathbb{Z}$ & $0$ \\
$\text{DIII}$ & $-1$ & $1$ & $1$ & $0$ & $\mathbb{Z}_2$ & $\mathbb{Z}_2$ & $\mathbb{Z}$ & $0$ & $0$ & $0$ & $\mathbb{Z}$ \\
$\text{AII}$ & $-1$ & $0$ & $0$ & $\mathbb{Z}$ & $0$ & $\mathbb{Z}_2$ & $\mathbb{Z}_2$ & $\mathbb{Z}$ & $0$ & $0$ & $0$ \\
$\text{CII}$ & $-1$ & $-1$ & $1$ & $0$ & $\mathbb{Z}$ & $0$ & $\mathbb{Z}_2$ & $\mathbb{Z}_2$ & $\mathbb{Z}$ & $0$ & $0$ \\
$\text{C}$ & $0$ & $-1$ & $0$ & $0$ & $0$ & $\mathbb{Z}$ & $0$ & $\mathbb{Z}_2$ & $\mathbb{Z}_2$ & $\mathbb{Z}$ & $0$ \\
$\text{CI}$ & $1$ & $-1$ & $1$ & $0$ & $0$ & $0$ & $\mathbb{Z}$ & $0$ & $\mathbb{Z}_2$ & $\mathbb{Z}_2$ & $\mathbb{Z}$ \\ \hline \hline

\end{tabular}}
\quad\\
\quad\\
\quad}

Here we rename the symmetric spaces with their Cartan labels, $1$ and $-1$ denotes the squares of the existent symmetries and $0$ denotes the absence of the symmetries. The columns correspond to the spatial dimension of the defined free-fermion Hamiltonian and $\mathbb{Z}$ and $\mathbb{Z}_2$ groups give the non-trivial topological phases in the relevant dimensions and symmetry classes.

\section{Complex Clifford bundles and complex classes}

In this section, by considering the vector bundles whose fibers corresponding to complex Clifford algebras, we will obtain the classification of complex classes of  topological phases.

Let us start with $\mathbb{C}l_k$-Dirac bundles. Since the complex Clifford algebras have two-fold periodicity as stated in (16), it is enough to consider $\mathbb{C}l_{k (\text{mod }2)}$. So, we can transform the $2\times 2$ table of complex Clifford algebras to the table of $\mathbb{C}l_{s-n(\text{mod }2)}$-Dirac bundles with $s,n=0,1$

\quad\\
{\centering{
\begin{tabular}{c c c}

% after \\: \hline or \cline{col1-col2} ...
\hline \hline
$\mathbb{C}l_{s-n (\text{mod }2)}$ & $n=0$ & $1$ \\ \hline
$s=0$ & $\mathbb{C}l_0$ & $\mathbb{C}l_1$ \\
$1$ & $\mathbb{C}l_1$ & $\mathbb{C}l_0$ \\ \hline \hline

\end{tabular}}
\quad\\
\quad\\
\quad\\}

The pattern repeats itself for bigger values of $s$ and $n$ because of the two-fold periodicity. Similar to the real case, we can define $\mathbb{C}l_k$-Dirac operators $\displaystyle{\not}\mathbb{D}_k$ on those bundles and construct the following table

\quad\\
{\centering{
\begin{tabular}{c c c}

% after \\: \hline or \cline{col1-col2} ...
\hline \hline
$\displaystyle{\not}\mathbb{D}_{s-n (\text{mod }2)}$ & $n=0$ & $1$ \\ \hline
$s=0$ & $\displaystyle{\not}\mathbb{D}_0$ & $\displaystyle{\not}\mathbb{D}_1$ \\
$1$ & $\displaystyle{\not}\mathbb{D}_1$ & $\displaystyle{\not}\mathbb{D}_0$ \\ \hline \hline

\end{tabular}}
\quad\\
\quad\\
\quad\\}

The index theorem for $\mathbb{C}l_k$-Dirac operators relates the analytic index of $\displaystyle{\not}\mathbb{D}_k$ with the topological invariants of the manifold. We have the following equality for the complex case
\begin{eqnarray}
\text{ind}(\displaystyle{\not}\mathbb{D}_k)&=&\left\{
                                                                               \begin{array}{ll}
                                                                                 \text{Td}(M), & \hbox{ for $k \text{ even}$ } \\
                                                                                 0, & \hbox{ for $k \text{  odd}$}                                                                                 
                                                                               \end{array}
                                                                             \right.
\end{eqnarray}
where $\text{Td}(M)$ denotes the Todd class and it is defined in terms of a power series expansion as
\begin{equation}
\text{Td}(M)=\prod_{i=1}^n\frac{x_i}{1-e^{-x_i}}
\end{equation}
It can be written in terms of the Chern classes $c_i$ as
\begin{equation}
\text{Td}(M)=1+\frac{1}{2}c_1+\frac{1}{12}\left(c_1^2+c_2\right)+\frac{1}{24}c_1c_2+...
\end{equation}
Todd class is an integer number and the index of $\displaystyle{\not}\mathbb{D}_k$ takes integer values for $k$ even and 0 otherwise. Then, we can write the table of $\mathbb{C}l_k$-Dirac operators in terms of the index of $\displaystyle{\not}\mathbb{D}_k$ as follows

\quad\\
{\centering{
\begin{tabular}{c c c}

% after \\: \hline or \cline{col1-col2} ...
\hline \hline
$\text{ind}\left(\displaystyle{\not}\mathbb{D}_{s-n (\text{mod }2)}\right)$ & $n=0$ & $1$ \\ \hline
$s=0$ & $\text{Td}(M)$ & $0$ \\
$1$ & $0$ & $\text{Td}(M)$ \\ \hline \hline

\end{tabular}}
\quad\\
\quad\\
\quad\\}

The Dirac operator $\displaystyle{\not}\mathbb{D}$ on a complex Dirac bundle is an element of the set of complex Fredholm operators $\mathfrak{F}_{\mathbb{C}}$ and $\mathfrak{F}_{\mathbb{C}}$ is a classifying space for $K$-groups of complex vector bundles. For a Haussdorf space $X$, index defines an isomorphism
\[
\text{ind}:[X,\mathfrak{F}_{\mathbb{C}}]\longrightarrow K(X)
\]
Then, we have the isomorphism between the homotopy groups of $\mathfrak{F}_{\mathbb{C}}$ and the $K$-groups at the point
\[
\text{ind}:\pi_k(\mathfrak{F}_{\mathbb{C}})\longrightarrow K^{-k}(\text{pt})\equiv \widetilde{K}(S^k)
\]
By similar reasoning as in the real case, the set $\mathfrak{F}_k$ of Fredholm operators which are $\mathbb{Z}_2$-graded, $\mathbb{C}l_k$-linear and self-adjoint is a classifying space for $K^{-k}$ and we have the isomorphism
\[
\text{ind}:[X,\mathfrak{F}_k]\longrightarrow K^{-k}(X)
\]
So, the index of $\mathbb{C}l_k$-Dirac operators $\displaystyle{\not}\mathbb{D}_k$ takes values in the group $K^{-k}(\text{pt})$ and we have the table

\quad\\
{\centering{
\begin{tabular}{c c c}

% after \\: \hline or \cline{col1-col2} ...
\hline \hline
$K^{-(s-n) (\text{mod }2)}(\text{pt})$ & $n=0$ & $1$ \\ \hline
$s=0$ & $K(\text{pt})$ & $K^{-1}(\text{pt})$ \\
$1$ & $K^{-1}(\text{pt})$ & $K(\text{pt})$ \\ \hline \hline

\end{tabular}}
\quad\\
\quad\\
\quad\\}

where the $K$-groups of the point are given as
\[
K(\text{pt})=\mathbb{Z}\quad\quad,\quad\quad K^{-1}(\text{pt})=0
\]
and the table transforms into

\quad\\
{\centering{
\begin{tabular}{c c c}

% after \\: \hline or \cline{col1-col2} ...
\hline \hline
$K^{-(s-n) (\text{mod }2)}(\text{pt})$ & $n=0$ & $1$ \\ \hline
$s=0$ & $\mathbb{Z}$ & $0$ \\
$1$ & $0$ & $\mathbb{Z}$ \\ \hline \hline

\end{tabular}}
\quad\\
\quad\\
\quad\\}

ABS isomorphism relates the $K$-groups with the complex representations of Clifford algebras similar to the real case. We denote the group of equivalence classes of irreducible complex representations of $\mathbb{C}l_k$ as $\mathfrak{M}^{\mathbb{C}}_k$. For different values of $k$, the groups are given as

\quad\\
{\centering{
\begin{tabular}{c c c}

% after \\: \hline or \cline{col1-col2} ...
\hline \hline
$k$ & $0$ & $1$ \\ \hline
$\mathbb{C}l_k$ & \,\,\,\,$\mathbb{C}$\,\,\,\, & \,\,\,\,$\mathbb{C}\oplus\mathbb{C}$ \\
$\mathfrak{M}^{\mathbb{C}}_k$ & \,\,\,\,$\mathbb{Z}$\,\,\,\, & \,\,\,\,$\mathbb{Z}\oplus\mathbb{Z}$ \\ \hline \hline

\end{tabular}}
\quad\\
\quad\\
\quad\\}

and for $\widehat{\mathfrak{M}}^{\mathbb{C}}_k=\mathfrak{M}^{\mathbb{C}}_{k-1}$, the quotient groups correspond to

\begin{eqnarray}
\widehat{\mathfrak{M}}^{\mathbb{C}}_k/i^*\widehat{\mathfrak{M}}^{\mathbb{C}}_{k+1}&\cong&\left\{
                                                                               \begin{array}{ll}
                                                                                 \mathbb{Z}, & \hbox{ for $k$ \text{even}} \\
                                                                                 0, & \hbox{ for $k$ \text{odd}}
                                                                               \end{array}
                                                                             \right.
\end{eqnarray}
ABS isomorphism defines an isomorphism between the $K$-groups at the point and the quotients $\widehat{\mathfrak{M}}^{\mathbb{C}}_*/i^*\widehat{\mathfrak{M}}^{\mathbb{C}}_{*+1}=\bigoplus_{k\geq 0}\widehat{\mathfrak{M}}^{\mathbb{C}}_k/i^*\widehat{\mathfrak{M}}^{\mathbb{C}}_{k+1}$ as follows
\[
\phi:\widehat{\mathfrak{M}}^{\mathbb{C}}_*/i^*\widehat{\mathfrak{M}}^{\mathbb{C}}_{*+1}\longrightarrow K^{-*}(\text{pt})
\]
So, we can write the table of $K$-groups as

\quad\\
{\centering{
\begin{tabular}{c c c}

% after \\: \hline or \cline{col1-col2} ...
\hline \hline
$\widehat{\mathfrak{M}}^{\mathbb{C}}_{s-n(\text{mod }2)}/i^*\widehat{\mathfrak{M}}^{\mathbb{C}}_{s-n+1(\text{mod }2)}$ & $n=0$ & $1$ \\ \hline
$s=0$ & $\mathbb{Z}$ & $0$ \\
$1$ & $0$ & $\mathbb{Z}$ \\ \hline \hline

\end{tabular}}
\quad\\
\quad\\
\quad\\}

The quotients of complex Grothendieck groups are also related to the Cartan symmetric spaces. By considering the following groups
\[
U(2n)\longrightarrow U(n)\times U(n)\longrightarrow U(n)
\]
one can see that the quotients of them correspond to Cartan symmetric spaces and the connected components of them are given as follows

\quad\\
{\centering{
\begin{tabular}{c c c}

% after \\: \hline or \cline{col1-col2}
\hline \hline
$C_i$ & $\text{symmetric spaces}$ & $\pi_0(C_i)$ \\ \hline
$C_0$ & $U(2n)/U(n)\times U(n)$ & $\mathbb{Z}$ \\
$C_1$ & $U(n)\times U(n)/U(n)$ & $0$ \\ \hline \hline

\end{tabular}}
\quad\\
\quad\\
\quad\\}

These symmetric spaces are classifying spaces of complex vector bundles and we have the following isomorphism for a Haussdorf space $X$
\[
K^{-k}(X)\cong [X,C_k]
\]
and
\[
K^{-k}(\text{pt})\cong\pi_0(C_k)
\]
so, from the ABS isomorphism
\[
\widehat{\mathfrak{M}}^{\mathbb{C}}_k/i^*\widehat{\mathfrak{M}}^{\mathbb{C}}_{k+1}\cong\pi_0(C_k)
\]
we can write the table of Grothendieck groups in terms of the connected componenets of the symmetric spaces

\quad\\
{\centering{
\begin{tabular}{c c c}

% after \\: \hline or \cline{col1-col2} ...
\hline \hline
$\pi_0(C_{s-n(\text{mod }2)})$ & $n=0$ & $1$ \\ \hline
$s=0$ & $\pi_0(C_0)$ & $\pi_0(C_1)$ \\
$1$ & $\pi_0(C_1)$ & $\pi_0(C_0)$ \\ \hline \hline

\end{tabular}}
\quad\\
\quad\\
\quad\\}

From the property $\pi_0(C_{s-n(\text{mod }2)})=\pi_{2-n(\text{mod }2)}(C_s)$, the table is written as

\quad\\
{\centering{
\begin{tabular}{c c c}

% after \\: \hline or \cline{col1-col2} ...
\hline \hline
$\pi_{2-n(\text{mod }2)}(C_s)$ & $n=0$ & $1$ \\ \hline
$s=0$ & $\mathbb{Z}$ & $0$ \\
$1$ & $0$ & $\mathbb{Z}$ \\ \hline \hline

\end{tabular}}
\quad\\
\quad\\
\quad\\}

and the rows correspond to Cartan symmetric spaces and the columns are the $(2-n)$th homotopy group of the symmetric space. As in the real case, the relations between symmetric spaces and complex Cliffrod algebra extensions given in the following table can be used in the transformation of the table of homotopy groups

\quad\\
{\centering{
\begin{tabular}{c c c}

% after \\: \hline or \cline{col1-col2}
\hline \hline
$C_i$ & $$ & $\text{Extension}$ \\ \hline
$C_0$ & $$ & $\mathbb{C}l_0\longrightarrow\mathbb{C}l_1$ \\
$C_1$ & $$ & $\mathbb{C}l_1\longrightarrow\mathbb{C}l_2$ \\ \hline \hline

\end{tabular}}
\quad\\
\quad\\
\quad\\}

Then, the symmetric space $C_0$ corresponds to adding the free-fermion Hamiltonian $H$ with no extra symmetries and $C_1$ is the symmetric space of the extension of $H$ with a chiral symmetry $S$. These Clifford algebra extensions are equivalent to A and AIII symmetry classes in the Altland-Zirnbauer classification and we have the periodic table of topological insulators for complex classes

\quad\\
{\centering{
\begin{tabular}{c| c c c|c c}

% after \\: \hline or \cline{col1-col2} ...
\hline \hline
$\text{label}$ & $T$ & $C$ & $S$ & $0$ & $1$ \\ \hline
$\text{A}$ & $0$ & $0$ & $0$ & $\mathbb{Z}$ & $0$ \\
$\text{AIII}$ & $0$ & $0$ & $1$ & $0$ & $\mathbb{Z}$ \\ \hline \hline

\end{tabular}}
\quad\\
\quad\\
\quad}

This completes the derivation of the periodic table of complex classes from the Clifford chessboard and index theorems. Because of the two-fold periodicity of complex Clifford algebras, we can extend the above table periodically and obtain the full periodic table of topological insulators and superconductors by combining with the real classes as follows

\quad\\
{\centering{
\begin{tabular}{c| c c c|c c c c c c c c}

% after \\: \hline or \cline{col1-col2} ...
\hline \hline
$\text{label}$ & $T$ & $C$ & $S$ & $0$ & $1$ & $2$ & $3$ & $4$ & $5$ & $6$ & $7$ \\ \hline
$\text{A}$ & $0$ & $0$ & $0$ & $\mathbb{Z}$ & $0$ & $\mathbb{Z}$ & $0$ & $\mathbb{Z}$ & $0$ & $\mathbb{Z}$ & $0$ \\
$\text{AIII}$ & $0$ & $0$ & $1$ & $0$ & $\mathbb{Z}$ & $0$ & $\mathbb{Z}$ & $0$ & $\mathbb{Z}$ & $0$ & $\mathbb{Z}$ \\ \hline
$\text{AI}$ & $1$ & $0$ & $0$ & $\mathbb{Z}$ & $0$ & $0$ & $0$ & $\mathbb{Z}$ & $0$ & $\mathbb{Z}_2$ & $\mathbb{Z}_2$ \\
$\text{BDI}$ & $1$ & $1$ & $1$ & $\mathbb{Z}_2$ & $\mathbb{Z}$ & $0$ & $0$ & $0$ & $\mathbb{Z}$ & $0$ & $\mathbb{Z}_2$ \\
$\text{D}$ & $0$ & $1$ & $0$ & $\mathbb{Z}_2$ & $\mathbb{Z}_2$ & $\mathbb{Z}$ & $0$ & $0$ & $0$ & $\mathbb{Z}$ & $0$ \\
$\text{DIII}$ & $-1$ & $1$ & $1$ & $0$ & $\mathbb{Z}_2$ & $\mathbb{Z}_2$ & $\mathbb{Z}$ & $0$ & $0$ & $0$ & $\mathbb{Z}$ \\
$\text{AII}$ & $-1$ & $0$ & $0$ & $\mathbb{Z}$ & $0$ & $\mathbb{Z}_2$ & $\mathbb{Z}_2$ & $\mathbb{Z}$ & $0$ & $0$ & $0$ \\
$\text{CII}$ & $-1$ & $-1$ & $1$ & $0$ & $\mathbb{Z}$ & $0$ & $\mathbb{Z}_2$ & $\mathbb{Z}_2$ & $\mathbb{Z}$ & $0$ & $0$ \\
$\text{C}$ & $0$ & $-1$ & $0$ & $0$ & $0$ & $\mathbb{Z}$ & $0$ & $\mathbb{Z}_2$ & $\mathbb{Z}_2$ & $\mathbb{Z}$ & $0$ \\
$\text{CI}$ & $1$ & $-1$ & $1$ & $0$ & $0$ & $0$ & $\mathbb{Z}$ & $0$ & $\mathbb{Z}_2$ & $\mathbb{Z}_2$ & $\mathbb{Z}$ \\ \hline \hline

\end{tabular}}
\quad\\
\quad\\
\quad}

\section{Conclusion}

Periodic table of topological insulators and superconductors is a direct result of the classification of Clifford algebras. We show that the topological classes in various dimensions can be obtained from the index theorems of Dirac operators defined on Clifford bundles. This result reveals the fact that the topological invariants defined for different topological classes are related to the topological index of relevant Dirac operators. $\mathbb{Z}$- and $\mathbb{Z}_2$-invariants such as Chern numbers, winding numbers, Kane-Mele and Chern-Simons invariants correspond to the topological index classes of Dirac operators in relevant dimensions and symmetry classes. So, the classification of topological phases of matter arises from algebra and topology.

The characterization of topological phases by index theorems gives also meaning to the connections between quantum field theory and gravitational anomalies and topological states of matter. Since the local gauge, gravitational and mixed anomalies are intimately related to the index theorems, they can be used to construct analogies with the periodic table of topological matter \cite{Ryu Moore Ludwig, Furusaki et al}. Moreover, the approach given in the paper may also have implications on the case of topological phases with more unitary symmetries such as space symmetries. The relations between Clifford chessboard, index theorems and symmetry protected topological phases are worth to investigate for the complete classification of topological matter.

%\references%


\begin{references}

\bibitem{Haldane} Haldane F D M 1988 Model for a quantum Hall effect without Landau levels: condensed matter realization of the "parity anomaly" \emph{Phys. Rev. Lett.} \textbf{61} 2015

\bibitem{Kane Mele} Kane C L and Mele E J 2005 Quantum spin Hall effect in graphene \emph{Phys. Rev. Lett.} \textbf{95} 226801

\bibitem{Hasan Kane}  Hasan M Z and Kane C L 2010 Colloquium: Topological insulators \emph{Rev. Mod. Phys.} \textbf{82} 3045

\bibitem{Qi Zhang} Qi X-L and Zhang S-C 2011 Topological insulators and superconductors \emph{Rev. Mod. Phys.} \textbf{83} 1057

\bibitem{Bernevig Hughes Zhang} Bernevig B A, Hughes T L and Zhang S-C 2006 Quantum spin Hall effect and topological phase transition in HgTe quantum wells \emph{Science} \textbf{314} 1757

\bibitem{Konig et al} K\"{o}nig M, Wiedmann S, Br\"{u}ne C, Roth A, Buhmann H, Molenkamp L W, Q, X-L and Zhang S-C 2007 Quantum spin Hall insulator state in HgTe quantum wells \emph{Science} \textbf{318} 766

\bibitem{Kane Mele2}  Kane C L and Mele E J 2005 $\mathbb{Z}_2$ topological order and the quantum spin Hall effect \emph{Phys. Rev. Lett.} \textbf{95} 146802

\bibitem{Wang Qi Zhang} Wang Z, Qi X-L and Zhang S-C 2010 Equivalent topological invariants of topological insulators \emph{New J. Phys.} \textbf{12} 065007

\bibitem{Kitaev} Kitaev A 2009 Periodic table for topological insulators and superconductors \emph{AIP Conf. Proc.} \textbf{1134} 22

\bibitem{Ryu Schnyder Furusaki Ludwig} Ryu S, Schnyder A P, Furusaki A and Ludwig A W W 2010 Topological insulators and superconductors: tenfold way and dimensional hierarchy \emph{New J. Phys.} \textbf{12} 065010

\bibitem{Stone Chiu Roy} Stone M, Chiu C K and Roy A 2011 Symmetries, dimensions and topological insulators: the mechanism behind the face of the Bott clock. \emph{J. Phys. A: Math. Theor.} \textbf{44} 0450001

\bibitem{Abramovici Kalugin} Abramovici G and Kalugin P 2012 Clifford modules and symmetries of topological insulators \emph{Int. J. Geom. Meth. Mod. Phys.} \textbf{9} 1250023

\bibitem{Freed Moore} Freed D and Moore G M 2013 Twisted equivariant matter \emph{Ann. Henri Poincar\'{e}} \textbf{14} 1927

\bibitem{Budich Trauzettel} Budich J C and Trauzettel B 2013 From the adiabatic theorem of quantum mechanics to topological states of matter \emph{Phys. Stat. Sol.} \textbf{7} 109

\bibitem{Kaufmann et al} Kaufmann R M, Li D and Wehefritz-Kaufmann B 2016 Notes on topological insulators \emph{Rev. Math. Phys.} \textbf{28} 1630003

\bibitem{Altland Zirnbauer} Altland A and Zirnbauer M 1997 Nonstandard symmetry classes in mesoscopic normal-superconducting hybrid structures \emph{Phys. Rev. B} \textbf{55} 1142

\bibitem{Atiyah Singer} Atiyah M F and Singer I M 1963 The index of elliptic operators on compact manifolds \emph{Bull. Am. Math. Soc.} \textbf{69} 422

\bibitem{Atiyah Singer2} Atiyah M F and Singer I M 1968 The index of elliptic operators I \emph{Ann. Math.} \textbf{87} 484

\bibitem{Atiyah} Atiyah M F 1968 Bott periodicity and the index of elliptic operators \emph{Quart. J. Math. Oxford} \textbf{19} 113

\bibitem{Morimoto Furusaki} Morimoto T and Furusaki A 2013 Topological classification with additional symmetries from Clifford algebras \emph{Phys. Rev. B} \textbf{88} 125129

\bibitem{Fruchart Carpentier} Fruchart M and Carpentier D 2013 An introduction to topological insulators \emph{Comp. Rend. Phys.} \textbf{14} 779

\bibitem{Shen Shan Lu} Shen S-Q, Shan W-Y and Lu H-Z 2011 Topological insulator and the Dirac equation \emph{Spin} \textbf{1} 33 

\bibitem{Fukui Shiozaki Fujiwara Fujimoto} Fukui T, Shiozaki K, Fujiwara T and Fujimoto S 2012 Bulk-edge correspondence for Chern topological phases: a viewpoint from a generalized index theorem \emph{J. Phys. Soc. Jap.} \textbf{81} 114602

\bibitem{Lawson Michelsohn} Lawson B H and Michelsohn M L 1989 \emph{Spin Geometry} (Princeton, NJ: Princeton University Press)

\bibitem{Benn Tucker} Benn I M and Tucker R W 1987 \emph{An Introduction to Spinors and Geometry with Applications in Physics} (Bristol: IOP Publishing)

\bibitem{Atiyah Bott Shapiro} Atiyah M F, Bott R and Shapiro A 1964 Clifford modules \emph{Topology} \textbf{3} 3

\bibitem{Ryu Moore Ludwig} Ryu S, Moore J E and Ludwig A W W 2012 Electromagnetic and gravitational responses and anomalies in topological insulators and superconductors \emph{Phys. Rev. B} \textbf{85} 045104

\bibitem{Furusaki et al} Furusaki A, Nagaosa N, Nomura K, Ryu S and Takayanagi T 2013 Electromagnetic and thermal responses in topological matter: Topological terms, quantum anomalies and D-branes \emph{Comp. Rend. Phys.} \textbf{14} 871

\end{references}
\end{document}